\documentclass{article}
\usepackage{dcase2019_techrep,amsmath,graphicx,url,times,booktabs, tabularx}
\usepackage{bbm}
\usepackage{multirow}
\usepackage{subfigure}

\title{Integrating the data augmentation scheme with various classifiers for acoustic scene modeling}

 \name{Hangting Chen$^{1,2}$,
       Zuozhen Liu$^{1,2}$,
       Zongming Liu$^{1,2}$, 
       Pengyuan Zhang$^{1,2}\sthanks{Pengyuan Zhang is the corresponding author.}$,
       Yonghong Yan$^{1,2,3}$
       }
 \address{	$^1$Key Laboratory of Speech Acoustics \& Content Understanding, Institute of Acoustics, CAS, China\\
 	$^2$University of Chinese Academy of Sciences, Beijing, China\\
 	$^3$Xinjiang Laboratory of Minority Speech and Language Information Processing,\\
 	Xinjiang Technical Institute of Physics and Chemistry, CAS, China
  }

\begin{document}

\ninept
\maketitle

\begin{sloppy}

\begin{abstract}
This technical report describes the IOA team's submission for TASK1A of DCASE2019 challenge. Our acoustic scene classification (ASC) system adopts a data augmentation scheme employing generative adversary networks. Two major classifiers, 1D deep convolutional neural network integrated with scalogram features and 2D fully convolutional neural network integrated with Mel filter bank features, are deployed in the scheme. Other approaches, such as adversary city adaptation, temporal module based on discrete cosine transform and hybrid architectures, have been developed for further fusion. The results of our experiments indicates that the final fusion systems A-D could achieve an accuracy higher than $85\%$ on the officially provided fold 1 evaluation dataset.
\end{abstract}


\vspace{-0.3cm}
\section{Introduction}
\label{sec:intro}
\vspace{-0.3cm}

Acoustic scene classification (ASC) aims to classify sounds into one of predefined classes \cite{mesaros2016tut}. Detection and Classification of Acoustic Scenes and Events (DCASE) challenges organized by IEEE Audio and Signal Processing (AASP) Technical Committee are one of the biggest competitions for ASC task. The large-scale dataset provided by DCASE2019 presents a difficult challenge for the system's fitting ability and generalization.

The report describes the details of IOA team's submission for TASK1A of DCASE2019. More concretely, data augmentation schemes based on generative neural networks (GAN) as well as two major classifiers improve the system's performance. We use two types of features, Mel filter bank feature (FBank) and scalogram extracted by wavelets, and two types of neural networks, 2D fully convolutional neural networks (FCNN) and 1D deep convolutional neural networks (DCNN). Other techniques, such as adversary domain adaptation, temporal module based on discrete cosine transform (DCT), hybrid neural network architectures, are developed for model ensemble. Under the official fold 1 evaluation setup, the final fusion systems could achieve above $85\%$ accuracy in the evaluation set.  

The remainder of this report is organized as follows. Section \ref{sec:data_aug} describes the scheme of data augmentation. Section \ref{sec:class_clf} details the features and architectures of classifiers. Section \ref{sec:fusion_method} presents our two methods for fusion. Section \ref{sec:exp} shows the details of experiments. Section \ref{sec:results} covers the results of classifiers and fusion systems and makes some discussion. Section \ref{sec:conclusion} concludes our work.

\vspace{-0.3cm}
\section{The data augmentation scheme}
\label{sec:data_aug}
\vspace{-0.3cm}

\begin{figure}[!htb]
	\centering
	\centerline{\includegraphics[width=0.7\columnwidth]{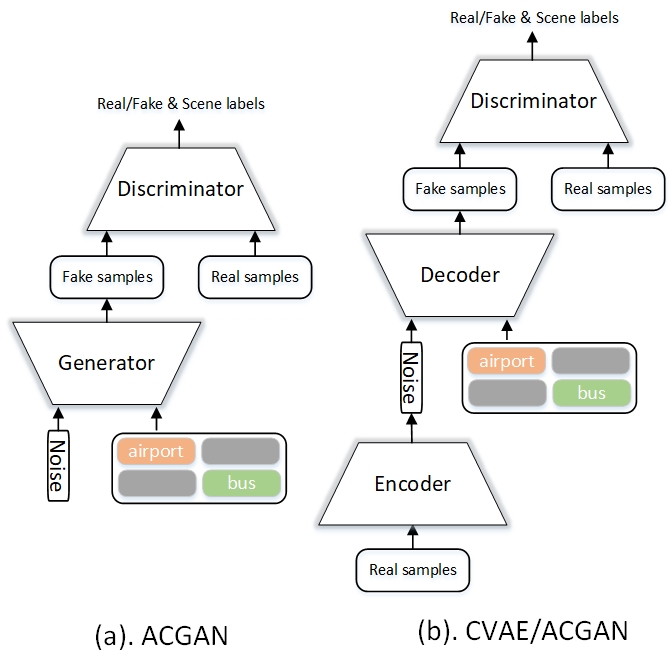}}
	\vspace{-0.5cm}
	\caption{(a) ACGAN and (b) CVAE/ACGAN architecture for data augmentation.}
	\label{fig:ganandvaegan}
\end{figure}
\vspace{-0.2cm}

\begin{figure}[!htb]
	\centering
	\centerline{\includegraphics[width=0.7\columnwidth]{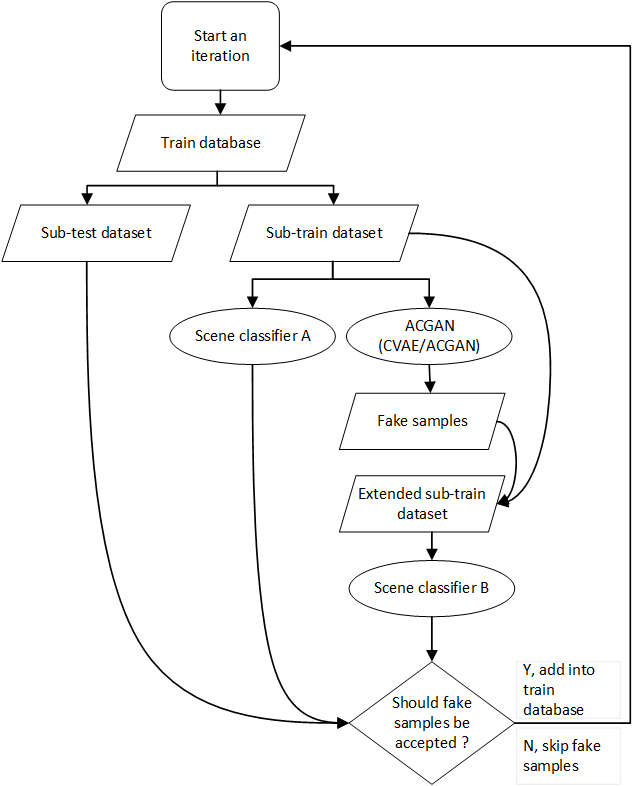}}
	\vspace{-0.5cm}
	\caption{The scheme of data augmentation.}
	\label{fig:scheme}
\end{figure}
\vspace{-0.2cm}

Though most ASC systems can accurately classify training samples, they suffer from inferring test records, especially those from unseen cities \cite{Mesaros2017}\cite{Mesaros2018}. To improve generalization, additional samples are generated and added into the database. Inspired by the rapid development of generative models in deep learning, auxiliary classifier GAN (ACGAN) \cite{odena2017conditional} are utilized to generate fake samples (Figure \ref{fig:ganandvaegan}(a)). The generator learns to create acoustic feature maps which look real with scene labels as an additional input condition. On the other hand, the discriminator learns to distinguish real features from fakes as well as scene labels. Thus the generator/discriminator aims to maximum/minimize the binary real/fake loss
\vspace{-0.15cm}
\begin{equation}
	\label{eqn:acgan_realfake}
	L_{real/fake}=\sum_{i} (log(Dis(x_i))+log(1-Dis(Gen(y_i,z)))),
\end{equation}
and to minimize the scene classification loss,
\vspace{-0.15cm}
\begin{equation}
	\label{eqn:acgan_label}
	L_{scene}=\sum_{i}\sum_{a\in A}{\mathbbm I}_{[a=y_i]}(log(Dis(x_i))+log(Dis(Gen(y_i,z)))),
\end{equation} 
where $x_i$,$y_i$,$z$ represents the networks' input, target and Gaussian noise, $A$ is a collection of scene classes, ${\mathbbm I}$ is the indicator function. The loss of ACGAN is defined as,
\vspace{-0.15cm}
\begin{equation}
	L_{ACGAN}=L_{real/fake}+\gamma L_{scene},
\end{equation}
where $\gamma$ controls the ratio between real/fake loss and scene classification loss.

The noise to generate fake features is usually assumed following Gaussian distribution, which actually makes little sense. To obtain samples as real as possible, a CVAE/ACGAN architecture  (Figure \ref{fig:ganandvaegan}(b)) is deployed as an alternative to ACGAN. It uses encoder to encode the real samples into noise restricted by a Kullback-Leibler (KL) loss,
\vspace{-0.15cm}
\begin{equation}
	L_{KL}=\sum_{i}D_{KL}(Enc(x_i)|| \mathcal{N}(0,I)),
\end{equation}
with respect to encoder with $x_i$ being the training sample. Besides, a reconstruction loss enables the reconstructed ones similar in contrast with the original input. In our framework, a mean-square loss of high-level bottleneck feature in the $l$th layer of the discriminator serves as reconstruction loss \cite{larsen2016autoencoding},
\vspace{-0.15cm}
\begin{equation}
	L_{reco}=\sum_{i} (Dis_l(x_i)-Dis_l(\tilde{x}_i))^2,
\end{equation}
where $\tilde{x}_i$ is the reconstruction of $x_i$. Therefore, the CVAE/ACGAN loss is defined as,
\vspace{-0.15cm}
\begin{equation}
	\begin{aligned}
		L_{CVAE/ACGAN}=&L_{real/fake}+\gamma_1 L_{scene} \\
		&+\gamma_2 L_{KL}+\gamma_3 L_{reco},
	\end{aligned}
\end{equation}
where $\gamma_1$,$\gamma_2$,$\gamma_3$ are the coefficients to balance various losses. 

The complete framework is plotted in Figure \ref{fig:scheme}, denoted as ACGAN or CVAE/ACGAN data augmentation scheme. In each iteration, the train database is firstly split into non-overlapped sub-train and sub-test set. Then a base classifier A is trained and tested on the sub-datasets. Also a generative model is trained on the sub-train set and then sampled when its output is stabilized. The fake samples are examined before added into the whole dataset. These generated candidates are mixed into sub-train set and another scene classifier B is trained. If its performance is improved on the sub-test dataset, these fake candidates will be accepted and added into the whole database.

Compared with \cite{Mun2017}, the fakes from different scenes can be sampled directly from the ACGAN (or CVAE/ACGAN) with scene labels as the condition. No need for training individual GANs for each scene class. Moreover, the discriminator with an auxiliary classifier ensures the generated samples not only look real but also belong to the target scene labels.

\vspace{-0.3cm}
\section{Classification Systems}
\label{sec:class_clf}
\vspace{-0.3cm}

\subsection{FBank-FCNN Classifier}
\label{subsec:fbank_cnn_clf}
\vspace{-0.2cm}

The FBank-FCNN network architecture is shown in Table \ref{table:FBank-FCNN}, similar to the one proposed in \cite{Dorfer2018}. It is a VGG \cite{simonyan2015very} style Network with 10 repeatedly stacked convolution layers containing small convolutional kernels. The common techniques in deep learning, such as batch normalization, dropout and Rectified Linear Units (ReLU), are used following the convolutional operations. The final classification part is designed as an $1\times1$ convolutional layer by decreasing the amount of channels to $10$ followed by a global average pooling layer over 10 feature maps, and finally a $10$-way SoftMax to the segment-level prediction.

\vspace{-0.4cm}
\begin{table}[!htb]
	\caption{The FCNN Classifier. The input feature map is of size frames($L$) $\times$ channels($c$) $\times$ filters($n$). The notation "$5 \times 5$ Conv(pad=$2$,stride=$2$)$\times 14c$-BN-ReLU" denotes a convolutional kernel with $14c$ output channels and a size of $5 \times 5$,followed by batch normalization and ReLU activation.}
	\label{table:FBank-FCNN}
	\centering
	\resizebox{0.48\textwidth}{!}{
		\begin{tabular}{cc}
			\hline  
			Layer Name & Settings \\
			\hline
			\hline
			Input & Fbank $L\times c\times n$ \\
			\hline
			\multirow{2}{2.0cm}{\centering{Conv1}} & $5 \times 5$ Conv(pad=$2$,stride=$2$)$\times 14c$-BN-ReLU \\
			& $3 \times 3$ Conv(pad=$1$,stride=$1$)$\times 14c$-BN-ReLU \\
			& $2 \times 2$ MaxPooling \\
			\hline
			\multirow{2}{2.0cm}{\centering{Conv2}} & $3 \times 3$ Conv(pad=$1$,stride=$1$)$\times 28c$-BN-ReLU    \\
			& $3 \times 3$ Conv(pad=$1$,stride=$1$)$\times 28c$-BN-ReLU    \\
			& $2 \times 2$ MaxPooling \\
			\hline
			\multirow{2}{2.0cm}{\centering{Conv3}} & $3 \times 3$ Conv(pad=$1$,stride=$1$)$\times 56c$-BN-ReLU    \\
			& Dropout($p=0.3$)  \\
			& $3 \times 3$ Conv(pad=$1$,stride=$1$)$\times 56c$-BN-ReLU    \\
			& Dropout($p=0.3$)  \\
			& $3 \times 3$ Conv(pad=$1$,stride=$1$)$\times 56c$-BN-ReLU    \\
			& Dropout($p=0.3$)  \\
			& $3 \times 3$ Conv(pad=$1$,stride=$1$)$\times 56c$-BN-ReLU    \\
			& $2 \times 2$ MaxPooling \\
			\hline
			\multirow{2}{2.0cm}{\centering{Conv4}} & $3 \times 3$ Conv(pad=$0$,stride=$1$)$\times 128c$-BN-ReLU    \\
			& Dropout($p=0.5$)  \\
			& $3 \times 3$ Conv(pad=$0$,stride=$1$)$\times 128c$-BN-ReLU    \\
			& Dropout($p=0.5$)  \\
			\hline
			\multirow{2}{2.0cm}{\centering{Pooling}} & $1 \times 1$ Conv(pad=$0$,stride=$1$)$\times 10$-BN-ReLU    \\
			& GlobalAveragePooling    \\
			\hline
			Output & 10-way SoftMax  \\
			\hline
	\end{tabular}}
\end{table}
\vspace{-0.2cm}

\subsection{Scalogram-DCNN Classifier}
\label{subsec:scalogram_cnn_clf}
\vspace{-0.2cm}

The Scalogram-DCNN classifier is based on \cite{chen2019audio}. The scalogram, which is used as an input of the DCNN classifier, is locally translation invariant and stable to time-warping deformation \cite{chen2018deep}. In this system, it is generated from wavelet filters operating on the spectrogram which is transformed from raw waveform. As shown in Table~\ref{tab:tab_DCNN}, the DCNN classifier consists of convolutional layers with small kernels and fully-connected (FC) layers.

\subsection{Other Classifiers}
\label{subsec:other_approach}
\vspace{-0.2cm}

\subsubsection{DCT Temporal Module}
\label{subsubsec:dct_temp_module}
\vspace{-0.2cm}

The scalogram-DCNN classifier is trained and evaluated in a frame-wise way. Due to the long term characteristics of wavelets, recurrent neural network can not achieve high classification accuracy. The temporal module based on DCT is deployed after the final affine transform as described in \cite{chen2019audio}. Different from the DCT filter in image processing, an attention weight filters the DCT spectrum by strengthening and weakening target feature map bins.

\subsubsection{Adversary City Adaptation}
\label{subsubsec:adv_city_training}
\vspace{-0.2cm}

To generalize the classifier for unseen cities, an adversary training branch, composed of a gradient reverse layer and a 2-layer feed-forward classifier, is connected following the convolutional layers in the scalogram-DCNN system. The branch classifies the record into the target city while a gradient reverse layer \cite{tzeng2017adversarial} makes the output of convolutional layers similar for the same scene class over various city domains. 

\vspace{-0.4cm}
\begin{table}[!htb]
	\caption{The DCNN Classifier. The input feature map is of size frames($L$) $\times$ channels($c$) $\times$ filters($n$). The notation 
		"$c\times3$ Conv(pad=$0$,stride=$1$)-$2c$-BN-ReLU" denotes a convolutional kernel with $c$ input channels, $2c$ output channels and a size of $3$, followed by batch normalization and ReLU activation.}
	\label{tab:tab_DCNN}
	\centering
	\resizebox{0.48\textwidth}{!}{
		\begin{tabular}{cc}
			\hline
			Layer Name & Settings \\
			\hline\hline
			Input & Scalogram $L\times c\times n$ \\
			\hline
			\multirow{2}{2.0cm}{\centering{Conv1}} & $c\times3$ Conv(pad=$0$,stride=$1$)$\times2c$-BN-ReLU             \\
			& $2$ Pooling(pad=$1$,stride=$2$)              \\
			\hline
			\multirow{2}{2.0cm}{\centering{Conv2}}  & $2c\times3$ Conv(pad=$0$,stride=$1$)$\times4c$-BN-ReLU             \\
			& $2$ Pooling(pad=$0$,stride=$2$)-Dropout              \\
			\hline
			\multirow{2}{2.0cm}{\centering{Conv3}} & $4c\times3$ Conv(pad=$0$,stride=$1$)$\times8c$-BN-ReLU             \\
			& $2$ Pooling(pad=$0$,stride=$2$)              \\
			\hline
			\multirow{2}{2.0cm}{\centering{Conv4}} & $8c\times3$ Conv(pad=$0$,stride=$1$)$\times16c$-BN-ReLU             \\
			& $2$ Pooling(pad=$0$,stride=$2$)-Dropout              \\
			\hline
			\multirow{2}{2.0cm}{} & Concatenate and flatten input as well as Conv's output \\ 
			\hline
			FC1 & Linear ($1024$ units)-BN-ReLU-Dropout \\ 
			\hline
			FC2 & Linear ($1024$ units)-BN-ReLU-Dropout \\ 
			\hline
			FC3 & Linear ($1024$ units)-BN-ReLU \\ 
			\hline
			Output & $10$-way SoftMax \\ 
			\hline
	\end{tabular}}
\end{table}
\vspace{-0.2cm}

\subsubsection{Hybrid Network Architecture}
\label{subsubsec:hybrid_net}
\vspace{-0.2cm}

\begin{figure}[!htb]
	\centering
	\centerline{\includegraphics[width=0.7\columnwidth]{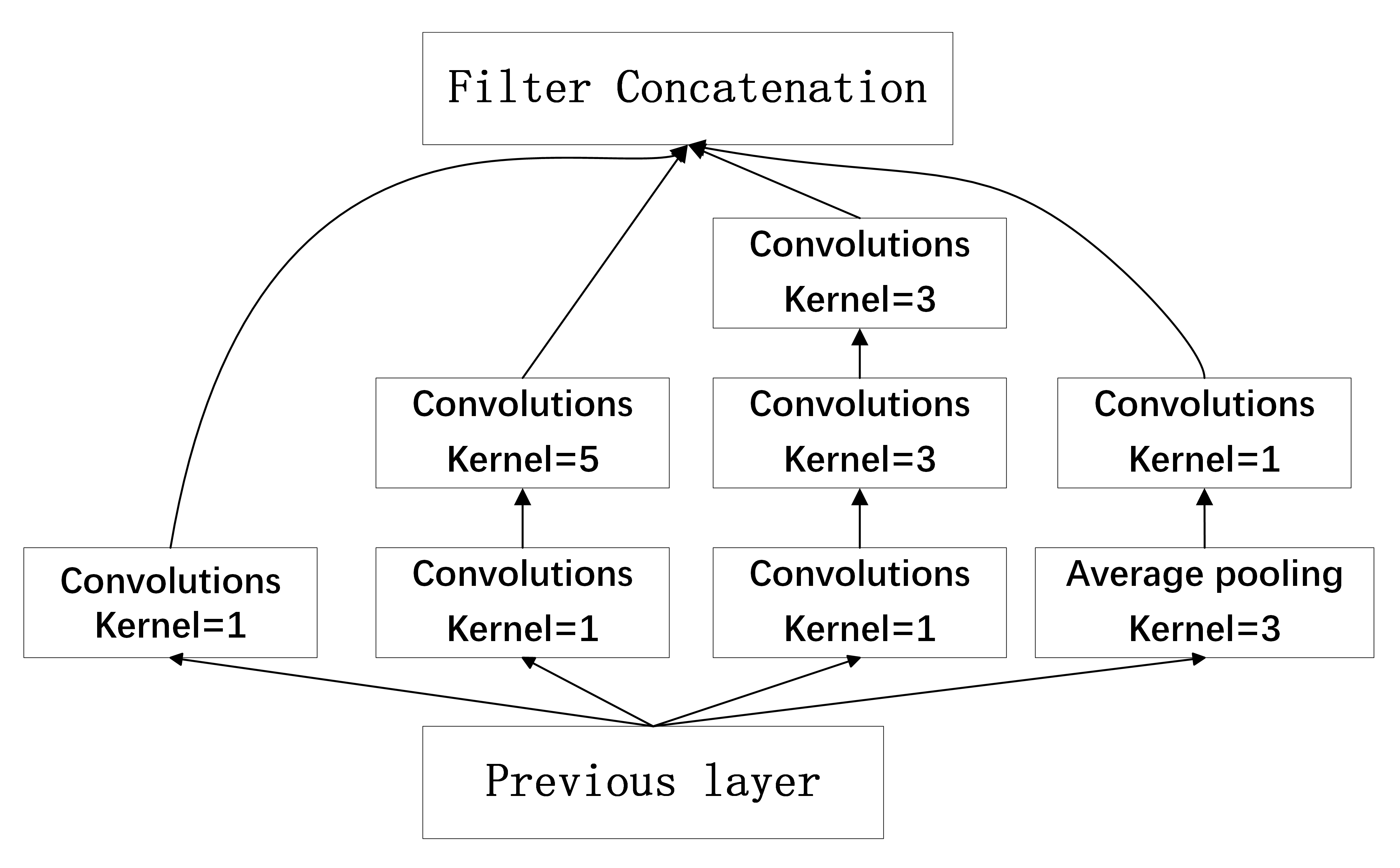}}
	\vspace{-0.5cm}
	\caption{Inception module I network architecture.}
	\label{fig:Inception1}
\end{figure}
\vspace{-0.2cm}

\begin{figure}[!htb]
	\centering
	\centerline{\includegraphics[width=0.7\columnwidth]{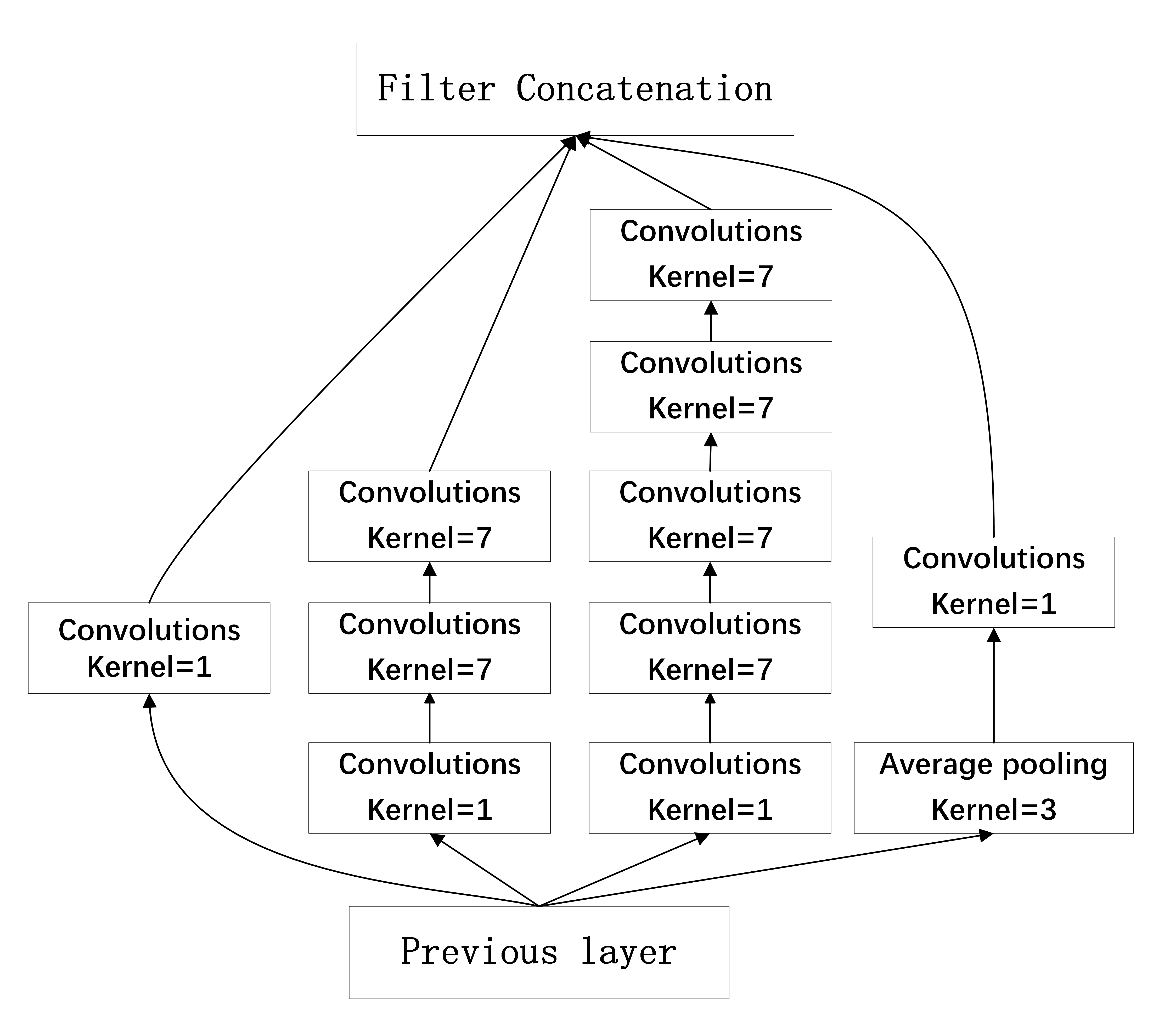}}
	\vspace{-0.5cm}
	\caption{Inception module II network architecture.}
	\label{fig:Inception2}
\end{figure}
\vspace{-0.2cm}

Several hybrid network architectures are proposed as classifiers. In many machine learning tasks using deep learning, increasing the size of networks can achieve better classification results. However, it may lead to a large amount of parameters which gives rise to the risk of overfitting due to limited labeled data. A deep convolutional network architecture codenamed Inception increased the depth and width of the network while keeping the computational budget constant \cite{szegedy2015going}. Then, improved versions of the Inception were proposed in \cite{szegedy2016rethinking}\cite{szegedy2017inception}. In order to expand the base network architectures with little parameter growth, we use two Inception modules (Figure~\ref{fig:Inception1} and Figure~\ref{fig:Inception2}) to replace the last 2 convolutional layers in the Scalogram-DCNN classifier. Unlike the work in the Inception networks, we try 1D and 2D CNN layers respectively in these Inception modules to maintain compatibility with the 1D CNN layers we use in the base DCNN architecture.

It is noticed that our DCNN network architecture with 1D convolutional layers may not make good use of temporal information. So recurrent layers like long short-term memory (LSTM) and gated recurrent unit (GRU) are added as a parallel channel for FC layers. Specifically, these layers use the output of the final convolutional layer as input. And their output as well as that of FC layers are combined as input to the last layers. 

After combining the Inception modules and recurrent layers into the base DCNN network, we get several hybrid network architectures, named IncepLSTM and IncepGRU. The IncepLSTM classifier uses $2$ layers of Inception module I instead of simple convolution and 2 layers of LSTM added as the parallel channel. The IncepGRUV1 classifier use 2 layers of Inception module II and 2 layers of GRU as the parallel channel. The IncepGRUV2 reduces the number of Inception module I to 1 followed with 1 layer of Inception module II. The IncepGRUV3 subsystem changes the convolutional layers of the Inception modules in V1 to 2D convolution. They are all expected to have some complementarity with the base classifiers. 

\vspace{-0.3cm}
\section{Ensemble methodology}
\label{sec:fusion_method}
\vspace{-0.3cm}

Different classifiers may lead to divisions among some controversial samples. Model ensemble can stabilize and generalize our final results. In the practice, voting serves as a simple and effective method compared with support vector machine, regression, etc. Two strategies, average and weighted voting, are adopted. The latter's weight is trained on the fold 1 evaluation set. 

\vspace{-0.3cm}
\section{Experiments}
\label{sec:exp}
\vspace{-0.3cm}

We used the officially provided fold 1 procedure to evaluate our systems' performance. Then the systems were retrained on the whole development data for submission. The train set was firstly split into the train and validate set. The classifiers were trained on the train set in maximum $200$ epochs. The validation set determined the early stopping of the training, i.e., the training would be stopped if its loss failed to decrease in continuous 5 epochs. We used Adam optimizer and set $\beta_1$ and $\beta_2$ to $0.9$ and $0.999$. The initial learning rate was $10^{-3}$ and was decreased according to the loss on the validate set. To relieve the influence of model's initialization, each system was trained in 3 different initial seeds. External data was not used in all experiments.

\subsection{Data augmentation}
\label{subsec:exp_dataaug}
\vspace{-0.2cm}

In each iteration of ACGAN (or CVAE/ACGAN) scheme, the train set was divided into sub-train and sub-test set of approximately equal size according to their recording cities. A base classifier was trained before and after the fake samples were added. Only if a performance improvement was observed, the fake samples were accepted and added into the whole database. In each iteration, we trained the ACGAN (or CVAE/ACGAN) for nearly $50$ epochs and sampled sets of spectrum from models in different training epochs.

\subsection{FBank-FCNN experiments}
\label{subsec:exp_fbank}
\vspace{-0.2cm}

We built two systems of different types of input features under FBank-FCNN architecture, one with left and right channel features as the input, another with difference and sum channel input. For both features, STFT was applied on the signal every $20ms$ over $40ms$ hamming windows. The total number of filters was 128. The Fbank of a $10$-second stereo audio was of the dimension $500\times2\times128$ before and $500\times6\times768$ after adding delta and delta-delta coefficients. For data augmentation, the generator and the discriminator were simplified versions of the classifier in Section \ref{subsec:fbank_cnn_clf}.

\subsection{Scalogram-DCNN experiments}
\label{subsec:exp_wavelet}
\vspace{-0.2cm}

To extract the scalogram, STFT was applied on the raw signal every $185ms$ over $555ms$ windows. The total number of wavelet filters was set to $290$, distributed uniformly at low frequency and logarithm at high frequency as described in \cite{Joakim2013Deep}. The scalogram of a 10-second stereo record was in a dimension of $58\times2\times290$. We followed the detailed settings of wavelets in \cite{chen2019audio}. The segment-wise prediction was obtained by accumulating the frame-wise output from the scalogram-DCNN classifier. Also the generator and discriminator adopted a simplified version of the classifier in Section \ref{subsec:scalogram_cnn_clf}.

\vspace{-0.3cm}
\section{Results and Discussion}
\label{sec:results}
\vspace{-0.3cm}

\vspace{-0.4cm}
\begin{table}[!htb]
	\caption{Results of experiments of different data augmentation frameworks on the fold 1 evaluation set, where the best performance is in bold.}
	\label{Tab:data_aug}
	\centering
	\resizebox{0.47\textwidth}{!}{
		\begin{tabular}{cccc}
			\hline
			\textbf{Feature type} & \textbf{Channels} & \textbf{Data augmentation scheme} & \textbf{Accuracy(\%)} \\
			\hline\hline
			Fbank & Left-Right & w/o & $76.92$ \\
			\hline
			Fbank & Left-Right & ACGAN & $\mathbf{77.56}$ \\
			\hline
			FBank & Ave-Diff & w/o & $79.95$ \\
			\hline
			FBank & Ave-Diff & ACGAN & $\mathbf{80.10}$ \\
			\hline\hline
			Scalogram & Left-Dight & w/o & $77.03$ \\
			\hline
			Scalogram & Left-Dight & ACGAN & $\mathbf{80.98}$ \\
			\hline
			Scalogram & Ave-Diff & w/o & $82.28$ \\
			\hline
			Scalogram & Ave-Diff & ACGAN & $84.06$ \\
			\hline
			Scalogram & Ave-Diff & CVAE/ACGAN & $\mathbf{84.28}$ \\
			\hline
	\end{tabular}}
\end{table}
\vspace{-0.2cm}

In this section, the complete results of the fold 1 evaluation setup on different schemes and classifiers are reported. The feature extracted from left-right and ave-diff (average-difference) channels was evaluated under FBank-FCNN and scalogram-DCNN classifiers with and without data augmentation. As listed in Table \ref{Tab:data_aug}, the features extracted from ave-diff channels could outperform that from left-right channels, approximately $3\%$-$5\%$. In addition, the GAN scheme can improve all classifiers performance from $0.5\%$ to $4\%$. The CVAE/ACGAN scheme could further give rise to higher accuracy in the scalogram-DCNN.

Two strategies were deployed in our following experiments. First, the ACGAN was adopted as a main scheme instead of CVAE/ACGAN, whose training was relatively slow. Moreover, ACGAN was already pretty distinct and when integrated with other techniques, it seemed more preferable to CVAE/ACGAN. Second, the Scalogram-DCNN classifiers trained only on the ave-diff features were served as a main force of system ensemble and FBank-FCNN classifiers trained both on the ave-diff and left-right features as supplementary. The results are listed in Table \ref{Tab:all_restults}, where we denote the name of each system as "\{feature type\}-\{feature channel\}-\{data augmentation scheme\}-\{classifiers\}". For example, "scalogram-avediff-ACGAN-city\_adversary" represents the system trained under ACGAN scheme using ave-diff scalogram and the adversary city adaptation classifier.

\vspace{-0.4cm}
\begin{table}[!htb]
	\caption{Results of experiments of various systems on the fold 1 evaluation set, where the top 3 classifiers are in bold.}
	\label{Tab:all_restults}
	\centering
	\resizebox{0.47\textwidth}{!}{
		\begin{tabular}{clc}
			\hline
			\textbf{Classifier ID} & \textbf{Classifier Name} & \textbf{Accuracy(\%)} \\
			\hline\hline
			1 & Fbank-leftright-ACGAN-FCNN & $77.56$ \\
			\hline
			2 & Fbank-avediff-ACGAN-FCNN & $80.10$ \\
			\hline\hline
			3 & Scalogram-avediff-ACGAN-DCNN & $84.06$ \\
			\hline
			4 & Scalogram-avediff-ACGAN-DCNN\_DCT & $83.58$ \\
			\hline
			5 & Scalogram-avediff-ACGAN-IncepLSTM & $83.80$ \\
			\hline
			6 & Scalogram-avediff-ACGAN-IncepGRU & $83.87$ \\
			\hline
			7 & Scalogram-avediff-ACGAN-city\_adversary & $\mathbf{84.16}$ \\
			\hline
			8 & Scalogram-avediff-ACGAN-city\_adversary\_DCT & $\mathbf{84.23}$ \\
			\hline
			9 & Scalogram-avediff-CVAE/ACGAN-DCNN & $\mathbf{84.28}$ \\
			\hline
			10 & Scalogram-avediff-CVAE/ACGAN-city\_adversary & $82.94$ \\
			\hline
	\end{tabular}}
\end{table}
\vspace{-0.2cm}

\vspace{-0.4cm}
\begin{table}[!htb]
	\caption{Results of fusion systems on the fold 1 evaluation set. System 2 and 3 used different weight.}
	\label{Tab:fusion_results}
	\centering
	\resizebox{0.47\textwidth}{!}{
		\begin{tabular}{clcc}
			\hline
			\textbf{System ID} & \textbf{Classifiers' IDs} & \textbf{Voting methods} & \textbf{Accuracy(\%)} \\
			\hline\hline
			A & 1,2,3,4,5,6,7 & Average & $85.07$ \\
			\hline
			B & 1,2,3,4,7,8 & Weighted & $85.11$ \\
			\hline
			C & 1,2,3,4,7,8 & Weighted & $85.11$ \\
			\hline
			D & 1,2,3,4,5,6,7,9 & Average & $\mathbf{85.28}$ \\
			\hline
	\end{tabular}}
\end{table}
\vspace{-0.2cm}

After data augmentation, the system built on wavelet features could always outperform the ones using FBank (1,2 and 3-10). The adversary city adaptation (3 and 7) and CVAE/ACGAN scheme (3 and 9) could lead to improvement but combing both failed to give better results (7,9,10). Additionally, the hybrid networks did not outperform the scalogram-avediff-ACGAN-DCNN ones (5,6 and 3). The DCT temporal module slightly promote the accuracy (7 and 8) but may harm the systems in some cases (3 and 4). The top $3$ classifiers were scalogram-avediff-CVAE/ACGAN-DCNN, scalogram-avediff-ACGAN-city\_adversary\_DCT and scalogram-avediff-ACGAN-city\_adversary, which all employed wavelet filters, average-difference channels' features and data augmentation scheme.

In fusion systems, FBank and scalogram features could be relatively complemented under a proper combination strategy. The detailed voting systems for the final submission are listed in Table \ref{Tab:fusion_results}, two using the weighted voting and two using the average voting.  

\vspace{-0.3cm}
\section{Conclusion}
\label{sec:conclusion}
\vspace{-0.3cm}

This report describes our submissions for DCASE2019 Task1A. In the fold 1 evaluation setup, the data augmentation scheme integrated with convolutional neural networks and other training architectures achieved accuracies above $77\%$ and $83\%$ for FBank and scalogram. After voting fusion, the final systems could achieve accuracies above $85\%$.

\bibliographystyle{IEEEtran}
\bibliography{refs}
%
%
%
%
%
%
%
%
%

\end{sloppy}
\end{document}